\documentstyle[12pt,epsf]{article}
\begin{document}
%
%
%
\newcommand{\al}{\mbox{$\alpha$}}
\newcommand{\B}{\mbox{$\beta$}}
\newcommand{\G}{\mbox{$\gamma$}}
\newcommand{\Fpi}{\mbox{$F_\pi$}}
\newcommand{\lb}[1]{\mbox{$\bar{l}_#1$}}
\newcommand{\pipi}{\mbox{$\pi\pi\mbox{ }$}}
\begin{flushright}  BUTP-96/10, hep-ph/9604217
\end{flushright}
\begin{center}
{\Large {\bf Scattering Lengths and
Medium and High Energy
\pipi Scattering}}
\vskip 2.5cm
{B.~Ananthanarayan\\
P.~B\"uttiker \\ 
Institut f\"ur theoretische Physik
\\Universit\"at Bern, 
CH--3012 Bern, Switzerland\\
}
\end{center}
%
%
\begin{abstract}
Medium and high energy absorptive parts
contribute to dispersive
expressions for D- wave scattering lengths, $a^0_2$ and $a^2_2$.
For the model employed by
Basdevant, Frogatt and Peterson we find the
D- wave driving term contributions to the D- wave scattering
lengths are $1.8\cdot 10^{-4}$ to $a^0_2$ and
$0.4\cdot 10^{-4}$ to $a^2_2$,  roughly
$10\%$ and $30\%$ of their respective
 central experimental values.
Inequivalent sets of sum rules are used as a compelling
test of the consistency of the model for which
crossing symmetry is not guaranteed.  
Results for the F- wave scattering length $a^1_3$ are presented,
completing the recent Roy equation analysis of \pipi scattering
in the range for $a^0_0$ favored by standard chiral perturbation theory.
\end{abstract}
\newpage
\setcounter{equation}{0}
\section{Introduction}
The Roy equations~\cite{roy,BGN},
a system of integral equations for physical region partial
wave amplitudes based on (fixed t-) dispersion relations, 
provide the necessary tools besides analyticity, unitarity
and crossing~\cite{martin,MMS}
for the analysis of low energy S- and P- wave scattering~\cite{BFP1,BFP2}.
Best fits to the experimental data and
Roy equations for the two lowest waves 
yield $a^0_0=0.26\pm 0.05$~\cite{nagels}.  
Chiral perturbation
theory, the low-energy limit of the
strong interactions~\cite{HL} among other things predicts
scattering lengths as well as higher threshold parameters of
\pipi scattering.  
Standard one-loop chiral perturbation theory  predicts,
for instance, $a^0_0=0.20
\pm 0.01$~\cite{g+l:ann}.  
A generalized version~\cite{KS} 
predicts larger values for $a^0_0$ and
is related to the question of small quark condensates in QCD; two-loop
\pipi scattering amplitudes in this framework were recently 
presented~\cite{gentwoloop}  in addition to a field-theoretic
calculation in the standard chiral perturbation theory~\cite{bijnens+}. 

A revival of interest in 
\pipi scattering and the underlying theory and 
assumptions has been recently witnessed~\cite{SFS}.
Furthermore, the method of Roy equation analysis 
of Basdevant, Frogatt and Petersen (BFP)~\cite{BFP1,BFP2}
has been  reimplemented in order to analyse \pipi scattering
data with $a^0_0$ chosen to lie in the range favored by standard chiral
perturbation theory~\cite{AB}. 
The end product of the BFP method is the availability of a parametric
representation for the absorptive
parts of the three lowest waves $f^0_0$, $f^2_0$ and $f^1_1$
in the region $4\leq s \leq 110$.  However
the Roy equations extend over the entire energy domain of
\pipi scattering and involves all partial waves:  the information
content of these is modeled in terms of the known resonances
such as what is now called
the $f_2(1270)$~\cite{pdg} (an $I=0$, $l=2$ state)
[but we will refer to this as $f_0$ in accordance with BFP] and in terms of
Regge phenomenology which then supply the {\it driving terms} to the
Roy equations of the S- and P- waves.  
BFP appeal to such a model and present
simple polynomial fits to the driving terms.  
The availability of the polynomial fits to
the S- and P- wave  driving terms yields sharp predictions for certain
(combinations) of S- and P- wave
threshold parameters that are correlated with the
choice of $a^0_0$ that was input in the recent 
Roy equation analysis~\cite{AB}.

Roy equations may also be written down for the higher waves and solved
in a manner discussed by  BFP and one may evaluate
the D-waves in the threshold region and obtain
the D- wave scattering lengths:  
however the availability of a parameteric representation
of the three lowest waves essentially contains all the information
required to compute these scattering lengths, with the higher wave
and high energy contributions coming from the appropriate limits
of the D- wave driving terms.  The latter are not available
in the literature  which prevented a sharp evaluation of these
scattering lengths that are correlated with $a^0_0$ in the range
favored by standard chiral perturbation theory~\cite{AB}.   

The purpose of this paper is to compute precisely such 
driving term contributions to the D- (and F-) wave
scattering lengths from the model of  BFP.
The D- and F- wave scattering lengths belong to a class
of (combinations of) threshold parameters that are crucial
in testing the predictions of chiral perturbation theory;
as accurate a determination of such quantities as possible is therefore
desirable.   Since experimental numbers for these are
in fact extracted from dispersion relation phenomenology,
it is important to have a handle on the relative contributions
of the low energy S- and P- waves and that of the medium and
high energy tails to the relevant dispersion integrals.

In the following we will recall the basis of the dispersion
relation analysis of Roy followed by a description of the BFP model,
 the details
of which in Ref.~\cite{BFP1,BFP2} are somewhat sketchy.
The first step therefore is to reconstruct the BFP model;
in order to establish the reconstruction we compute the S- and P- wave
driving terms from the model, 
obtained from the appropriate Roy dispersion relations for
amplitudes of definite isospin  projected
on these waves and compare them with the polynomial
fits provided by BFP~\cite{BFP1,BFP2}.   
Projecting on the D- (F-) wave, we obtain
the corresponding D- (F-) wave driving terms:  we will merely
evaluate these in the threshold region which will yield the
driving term contribution to the D- (F-) wave scattering lengths.

Indeed, it has been noted by BFP that crossing constraints are
not guaranteed to be satisfied by this model.  
In order to test the reliability of this determination, we then compute the
contributions of the medium and high energy information described
by the model to three sets of a priori
inequivalent sum rules for these scattering
lengths that are presently available in direct and indirect
forms the literature.
These are  obtained from considering (a) the Froissart-Gribov
representation for the D- waves in the threshold region~\cite{MMS,pp},
(b) those derived by Wanders~\cite{wan:66}, and (c) those derived
from a system of sum rules presented by Ananthanarayan, Toublan
and Wanders(ATW)~\cite{ATW}.
We continue with a presentation of numerical
details and a discussion of our results.   Implications
of this work  to the results of Ref.~\cite{AB} are discussed: in Ref.~\cite{AB}
only the resonance contributions of medium and high energy
information were accounted for which contributed $0.54 \cdot 10^{-4}$
($0.43 \cdot 10^{-4}$)
to $a^0_2$ and
$0.38\cdot 10^{-4}$ 
($0.31 \cdot 10^{-4}$)
to $a^2_2$ with
the Particle Data Group (BFP)
parameters for the $f_0$;  now the Regge and Pomeron contributions
may also be included which are $1.40\cdot 10^{-4}$ to 
$a^0_2$ and $0.07\cdot 10^{-4}$ to $a^2_2$ respectively.

In two-loop chiral perturbation theory, parameters
of the relevant effective lagrangian, enter the expressions for
the two-loop predictions to the F- wave scattering length $a^1_3$.
We employ the fits to the Roy equations
discussed in Ref.~\cite{AB} to compute the S- and P- wave contributions
to this important threshold parameter.  In practice $a^1_3$ receives
practically no contribution from the medium and high energy absorptive
parts; this is established
once more by computing the driving term contribution.
Sum rules for $a^1_3$  (a) obtained from the appropriate limit
of the Froissart-Gribov represenataion for the F- wave, and
(b) obtained by ATW~\cite{ATW2} are used to test the consistency
of the driving term contributions.  In practice, these are found
to be two orders of magnitude smaller than the contribution from
the S- and P- waves.

\setcounter{equation}{0}
\section{Dispersion relations and Roy equations}
The notation and formalism that
we adopt in this discussion follows that of Ref.~\cite{BFP2}.
Consider  \pipi scattering:  
\begin{eqnarray*}
	\pi^a (p_a) + \pi^b (p_b) \to \pi^c (p_c) + \pi^d (p_d),
\end{eqnarray*}
where all the pions have the same mass,
$m_\pi=140$ MeV and is henceforth set equal to unity.
[Unless explicitly mentioned all masses will be in the units of $m_\pi$.]
The Mandelstam variables $s$, $t$ and $u$ are
defined as
\begin{equation}
	s = (p_a + p_b)^2 ,\quad t = (p_a - p_c)^2 ,\quad t = (p_a - p_d)^2,%
	\quad s+t+u=4.
\end{equation}
The scattering amplitude $F(a,b \to c,d)$
(our normalization of the amplitude is
that of Ref.~\cite{BFP2}, and differs from that of Ref.~\cite{g+l:ann,AB}
by $32\pi$): 
\begin{eqnarray*}
  F(a,b \to c,d)  =  \delta_{ab} \delta_{cd} A(s,t,u) + \delta_{ac}
                     \delta_{bd} A(t,s,u) + \delta_{ad} \delta_{bc} A(u,t,s).
\end{eqnarray*}
From  $A(s,t,u)$ we construct the three $s$-channel isospin
amplitudes:
\begin{eqnarray}\label{eq:amp:iso:def}
T^{0}_s(s,t,u) & = & 3 A(s,t,u) + A(t,s,u) + A(u,t,s), \nonumber\\
T^{1}_s(s,t,u) & = & A(t,s,u) - A(u,t,s), \\
T^{2}_s(s,t,u) & = & A(t,s,u) + A(u,t,s). \nonumber
\end{eqnarray}
One basis for the dispersion relation analysis of \pipi scattering
data is dispersion relations for amplitudes of definite isospin
in the $s-$channel which may be written down with two subtractions
(a number that guarantees convergence as a result of the Froissart
bound, rigorously established in axiomatic field theory):
\begin{eqnarray}\label{dispersion}
T^I_s(s,t,u) & = &\sum_{I'=0}^2 C_{st}^{II'}(C^{I'}(t)+
(s-u)D^{I'}(t))+\nonumber \\
	   &   &{1\over\pi}\int_4^\infty {dx\over x^2}\left(
		{s^2\over x-s} {\bf I}^{II'} +
		{u^2\over x-u}C_{su}^{II'}\right)\ A^{I'}_s(x,t),
\end{eqnarray}
where $A^{I}_s(x,t)$ 
is the isospin $I$
$s-$channel absorptive part, $C_{st}$ and $C_{su}$ are the crossing matrices:
\begin{eqnarray*}
	C_{st} = \pmatrix{
	 1/3 & 1 & 5/3 \cr
	 1/3 & 1/2 & -5/6 \cr
	 1/3 & -1/2 & 1/6 \cr
	 },\quad
	C_{su} = \pmatrix{
	 1/3 & -1 & 5/3 \cr
	 -1/3 & 1/2 & 5/6 \cr
	 1/3 & 1/2 & 1/6 \cr
	 }\quad
%
\end{eqnarray*}
and ${\bf I}$ is the identity matrix.  Suppressing $u=4-s-t$ as
an argument of $T^I_s$,  
we introduce the partial wave expansion:
\begin{equation}
  T^I_s(s,t) = %
	\sum_{l=0}^{\infty} (2 l + 1) P_l(1+{2t\over s-4} ) f^I_l(s),
\end{equation}
\begin{eqnarray}\label{projection}
f^I_l(s)={1\over 4-s} \int_0^{4-s} dt
\, T^I_s(s,t) P_l(1+{2t\over s-4}), \nonumber 
\\
f^0_l(s)=f^2_l(s)=0, l\ {\rm odd},\quad f^1_l(s)=0, l\ {\rm even}.
\end{eqnarray}
In terms of the phase shifts, $\delta^I_l$, in the elastic region,
we have:
\begin{equation}
f^I_l(s)=\sqrt{\frac{s}{s-4}} \exp{i\delta^I_l(s)} \sin\delta^I_l(s),
\, 4\leq s\leq 16.
\end{equation}
We also introduce the threshold expansion:
\begin{equation}
{\rm Re}f^{I}_l(s)=\left({s-4\over 4}\right)^l\left(a^I_l+b^I_l
\left({s-4\over 4}\right)+\ldots\right), \, s>4,
\end{equation}
where the $a^I_l$ are the scattering lengths and the $b^I_l$ are
the effective ranges, namely the leading threshold parameters.

Roy eliminated the $t-$dependent unknown functions
$C^I(t)$ and $D^I(t)$ in eq.(\ref{dispersion})
using crossing symmetry and Bose symmetry
which implies: $C^1(t)=D^0(t)=D^2(t)=0$,
in favor of the S-wave
scattering lengths.  The result is the Roy form of
the fixed $t$ dispersion relations with two subtractions:
\begin{eqnarray}\label{roydispersion}
T^I_s(s,t)  & = & \sum_{I'=0}^2{1\over 4}
 g_1^{II'}(s,t) a^{I'}_0\nonumber \\
&  &+ \int_4^\infty  ds'\left[ g_2^{II'}(s,t,s') A_s^{I'}(s',0)
+ g_3^{II'}(s,t,s') A_s^{I'}(s',t) \right]
\end{eqnarray}
where the three functions $g_i^{II'}$  listed in eq. (15)-(17)
of Ref~\cite{roy} are listed below:
\begin{eqnarray*}
g_1(s,t) & = & s({\bf
I}-C_{su})+t(C_{st}-C_{su})
+4 C_{su}, \\
g_2(s,t,s') & = & C_{st}\left({{\bf I}+C_{tu}\over 2} +{2s+t-4\over t-4}
{{\bf I}-C_{tu}\over 2}\right){1\over \pi s'^2}\cdot \\
& & \left[{t^2{\bf I}\over s'-t} + {(4-t)^2 C_{su} \over s'-4+t} -
{4t{\bf I}+4(4-t)C_{su} \over s'-4} \right], \\
g_3(s,t,s')& = & {1\over \pi s'^2}\left[{s^2{\bf I}\over s'-s}+
{(4-s-t)^2 C_{su} \over s'-4+s+t}-{(4-t)^2\over s'-4+t} \right.\\
& & \left.\left\{ {C_{su}+{\bf I}\over 2} + {2s+t-4 \over t-4}{C_{su}-
{\bf I}\over 2}\right\}\right],
\end{eqnarray*}
with $C_{tu}^{II'}=(-1)^I \delta^{II'}$.
The Roy equations are obtained upon
projecting the resulting dispersion relation onto
partial waves and inserting a partial wave expansion for the absorptive part.
They have been rigorously proved to be valid in the domain
$4\leq s \leq 60$.
These are a system of coupled integral equations for
partial wave amplitudes of definite isospin $I$ which are related
through crossing symmetry to the absorptive parts of all the
partial waves.  
The Roy equations for the S- and P- waves are
\cite{roy,BGN,BFP1,BFP2}:
\begin{eqnarray}\label{roy_eq}
f^0_0(s) & = & a^0_0 + ( 2 a^0_0 - 5 a^2_0) \frac{s-4}{12}
      		+ \sum_{I'=0}^{2} \sum_{l'=0}^{\infty}\int_{4}^{\infty}dx
		K^{l'I'}_{00}(s,x) \mbox{Im} f^{I'}_{l'}(x),\nonumber\\
f^1_1(s) & = & ( 2 a^0_0 - 5 a^2_0) \frac{s-4}{72}
      		+ \sum_{I'=0}^{2} \sum_{l'=0}^{\infty}\int_{4}^{\infty}dx
		K^{l'I'}_{11}(s,x) \mbox{Im} f^{I'}_{l'}(x),\\
f^2_0(s) & = & a^2_0 - ( 2 a^0_0 - 5 a^2_0) \frac{s-4}{24}
      		+ \sum_{I'=0}^{2} \sum_{l'=0}^{\infty}\int_{4}^{\infty}dx
		K^{l'I'}_{20}(s,x) \mbox{Im} f^{I'}_{l'}(x) \nonumber
\end{eqnarray}
and for all the higher partial waves written as:
\begin{eqnarray*}
f^I_l(s) & = &  \sum_{I'=0}^{2} \sum_{l'=0}^{\infty}\int_{4}^{\infty}dx
		K^{l'I'}_{Il}(s,x) \mbox{Im} f^{I'}_{l'}(x),\quad l\geq 2,
\end{eqnarray*}
where $K^{l'I'}_{lI}(s,s')$ are 
the kernels of the integral equations and whose
explicit expressions have been documented
elsewhere~\cite{BGN}. 
Upon cutting off the integral at a large scale
$\Lambda$ 
and absorbing the contribution of the high energy tail as well as
that of all the higher waves over the entire energy range
into the driving terms $d^I_l(s,\Lambda)$ we have:
\begin{eqnarray}\label{roy_eq_drv}
f^0_0(s) & = & a^0_0 + ( 2 a^0_0 - 5 a^2_0) \frac{s-4}{12}
      		+ \sum_{I'=0}^{2} \sum_{l'=0}^{1}\int_{4}^{\Lambda}dx
		K^{l'I'}_{00}(s,x) \mbox{Im} f^{I'}_{l'}(x) \nonumber \\
& & +		d^0_0(s,\Lambda),\nonumber\\
f^1_1(s) & = & ( 2 a^0_0 - 5 a^2_0) \frac{s-4}{72}
      		+ \sum_{I'=0}^{2} \sum_{l'=0}^{1}\int_{4}^{\Lambda}dx
		K^{l'I'}_{11}(s,x) \mbox{Im} f^{I'}_{l'}(x) \nonumber \\
& & +		d^1_1(s,\Lambda),\\
f^2_0(s) & = & a^2_0 - ( 2 a^0_0 - 5 a^2_0) \frac{s-4}{24}
      		+ \sum_{I'=0}^{2} \sum_{l'=0}^{1}\int_{4}^{\Lambda}dx
		K^{l'I'}_{20}(s,x) \mbox{Im} f^{I'}_{l'}(x) \nonumber \\
& & +		d^2_0(s,\Lambda) \nonumber
\end{eqnarray}
and for all the higher partial waves written as:
\begin{eqnarray*}
f^I_l(s) & = &  \sum_{I'=0}^{2} \sum_{l'=0}^{1}\int_{4}^{\Lambda}dx
		K^{l'I'}_{Il}(s,x) \mbox{Im} f^{I'}_{l'}(x)
		+d^I_l(s,\Lambda),\quad l\geq 2.
\end{eqnarray*}
From the following limits for the Roy equations
\begin{eqnarray}
& \displaystyle \lim_{s\to 4+} {{\rm Re}f^0_2(s)\over((s-4)/4)^2},\quad
 \lim_{s\to 4+} {{\rm Re}f^2_2(s)\over ((s-4)/4)^2}, \quad
\lim_{s\to 4+} {{\rm Re}f^1_3(s)\over ((s-4)/4)^3} & 
\end{eqnarray}
we find expressions of sum rules for the D- and F- wave scattering lengths:
\begin{small}
\begin{eqnarray}
a^0_2  & = & \frac{16}{45\pi} \int^{\Lambda}_4 \frac{ds'}{s'^3(s'-4)}%
	        \nonumber \\
	  &   & \left\{(s'-4) {\rm Im}f^0_0(s') + 9 (s'+4) {\rm Im}f^1_1(s')%
		   + 5(s'-4) {\rm Im}f^2_0(s')\right\}  \\
	  &   & +\lim_{s\to 4+} \frac{{\rm Re\ } d^0_2(s,\Lambda)}
	{((s-4)/4)^2},
		 \nonumber \\
a^2_2  & = & \frac{16}{90\pi} \int^{\Lambda}_4 \frac{ds'}{s'^3(s'-4)}%
	        \nonumber \\
	  &   & \left\{2(s'-4) {\rm Im}f^0_0(s') - 9 (s'+4) {\rm Im}f^1_1(s')%
		   + (s'-4) {\rm Im}f^2_0(s')\right\}  \\
	  &   &  +\lim_{s\to 4+} 
\frac{{\rm Re\ }d^2_2(s,\Lambda)}{((s-4)/4)^2}
	,	\nonumber \\
a^1_3  & = & \frac{16}{105\pi} \int^{\Lambda}_4 \frac{ds'}{s'^4(s'-4)}%
	        \nonumber \\
	  &   & \left\{2(s'-4) {\rm Im}f^0_0(s') + 9 (s'+4) {\rm Im}f^1_1(s')%
		   -5 (s'-4) {\rm Im}f^2_0(s')\right\} \label{a13sp} \\
	  &   &  +\lim_{s\to 4+} 
\frac{{\rm Re\ }d^1_3(s,\Lambda)}{((s-4)/4)^3}
	.	\nonumber
\end{eqnarray}
\end{small}
The objects of interest to us here are the driving terms $d^I_l(s,\Lambda)$
and in particular the driving term contributions to the D- and F- wave
scattering lengths 
\begin{eqnarray*}
\lim_{s\to 4+} 
\frac{{\rm Re\ }d^I_2(s,\Lambda)}{((s-4)/4)^2},
\,
\lim_{s\to 4+} 
\frac{{\rm Re\ }d^1_3(s,\Lambda)}{((s-4)/4)^3}
\end{eqnarray*}
after the model
for the medium and high energy contributions is pinned down.
This will be discussed in the subsequent sections.

\setcounter{equation}{0}
\section{The BFP Model}

The medium and high energy absorptive parts are described by
BFP in terms of only one resonance, viz., the $f_0$ whose mass
is taken
to be $M_{f_0}=1269$ MeV and elastic width to be $\Gamma_{f_0}=125$ MeV.  
[More updated information on this resonance may be obtained 
from~\cite{pdg},viz.,
$M_{f_0}=1275$ MeV, $\Gamma=185$ MeV, with the \pipi branching ratio
of $85\%$, yielding an elastic width $\Gamma_{f_0}=158$ MeV.
We employ the BFP numbers since these have already gone into
the driving terms for the S- and P- waves in the implementation of
Ref.~\cite{AB}.]
While details of the exact implementation are not available, we
are faced with the option of representing this resonance in
terms of say, a modified Breit-Wigner propagator along the
lines of Pennington and Protopopescu~\cite{PennProt} or merely in the narrow
width approximation.  In practice we have found that the
contributions from the latter
when added to the contributions arising from the remainder of
the high energy model,
yields good agreement with the published polynomial fit
of BFP and we have chosen to work with it.
The expression for the absorptive part from the $f_0$ is
therefore given by
\begin{equation}\label{resonance}
A_s^0(s',t)=5 \pi \Gamma_{f_0} M_{f_0} \sqrt{{s'\over
s'-4}} \delta(s'-M_{f_0}^2)
P_2(1+2t/(s'-4)), 4\leq s' < \infty,
\end{equation}
and $A_s^1(s',t)=A_s^2(s',t)=0$.

BFP describe the high energy asymptotics in terms of (a) Pomeron
exchange, and (b) Regge trajectory due to an {\it exchange degenerate}
$\rho+f_0$ trajectory.  

(a) Pomeron exchange:  the BFP Pomeron is characterized by 
the logarithmic slope of the differential cross-section $b$, at 
an energy scale $x_0$, with 
the slope of the Pomeron trajectory
$\alpha_P'$, 
 and a total asymptotic cross-section,
$\sigma_\infty$.  In terms of these, 
our reconstruction of  the absorptive part in the
$I=0$, $t$-channel reads:
\begin{equation}\label{pomeron}
A_t^0(s',t)={3x_0\over 32\pi} \sigma_\infty e^{bt/2} ({s'\over x_0})^{1+
\alpha_P' t} \Theta(s'-110),
\end{equation}
while $A_t^1(s',t)=A_t^2(s',t)=0$, and $\Theta(z)$
is the step-function.   This is
obtained by suitably modifying the expressions for the absorptive
parts presented in Ref.~\cite{PennProt}.

The BFP Pomeron is defined by the numerical choice, $b=10$ GeV$^{-2}$,
$x_0=10$ GeV$^{2}$, $\alpha_P'=0.4$ GeV$^{-2}$, $\sigma_\infty=1/m_\pi^2=
20$ mb.

(b) Regge exchange:  the BFP Regge trajectories are exchange degenerate 
$\rho+f_0$ poles whose 
residues are described by the Lovelace-Veneziano function
with universal $\rho$ coupling 
\begin{eqnarray*}
{f^2\over 4 \pi}=2.4.
\end{eqnarray*}
BFP provide no further details of the Regge contributions.  In order
to reconstruct the above, we require the specification of 
the trajectory:
\begin{eqnarray}
\alpha (t)={1\over 2} + {t\over 2 M_\rho^2},
\end{eqnarray}
$M_\rho=769$ MeV.
We then have the absorptive parts in the $I=0,1$ $t$-channels:
\begin{eqnarray}\label{regge}
A_t^0(s',t)& = & A_t^1(s',t) = \nonumber \\
& &
{f^2\over 16\pi} \sin(\pi\alpha(t))\Gamma(1/2-\alpha(t))
\left({s'\over 2M_\rho^2}\right)^{\alpha(t)} \Theta(s'-110),
\end{eqnarray}
and $A_t^2(s',t)=0$.

In our numerical evaluation, we have chosen to work with
(a) retaining upto the next to leading order contribution in $\alpha_P'$
since it is numerically small, and
(b) the simplified Regge trajectory $\alpha(t)=1/2$ since the
Regge contributions are expected to be small and the retaining
the slope of the trajectory entails  higher order corrections in
$ M_\rho^{-2}$ which are small.

\setcounter{equation}{0}
\section{S- and P- wave driving terms and D- wave scattering lengths}

The relations eq.(\ref{resonance}),
(\ref{pomeron}) and (\ref{regge})
 completely specify the BFP medium and high energy
\pipi scattering model along with the crossing relation for the
$s-$ and $t-$
channel absorptive parts 
\begin{eqnarray}\label{crossingrelation}
A_s^I(s',t)=\sum_{I'=0}^2 C_{st}^{II'}A_t^{I'}(s',t),
\end{eqnarray}
and may be directly inserted into the dispersion 
relation~(\ref{roydispersion}) and then be subsequently projected
onto the relevant partial waves, via. eq.(\ref{projection}).  
The results of the projection on to the S- and P- waves 
may be compared with the polynomial fits
provided by BFP,~\cite{BFP1,BFP2}.  These are [BFP (I)]:
\begin{eqnarray}
d^0_0(s) & = &  9.12\cdot 10^{-4}(s-4) +  9.78\cdot 10^{-5}(s-4)^2  \\
d^1_1(s) & = &  3.00\cdot 10^{-5}(s-4) + 2.30\cdot 10^{-5}(s-4)^2 
 \nonumber \\
d^2_0(s) & = &  7.20\cdot 10^{-4}(s-4) + 3.50\cdot 10^{-5}(s-4)^2  \nonumber 
\end{eqnarray}
and [BFP (II)]:
\begin{eqnarray}
d^0_0(s) & = &  9.12\cdot 10^{-4}(s-4) +  9.78\cdot 10^{-5}(s-4)^2  \\
d^1_1(s) & = &  1.36\cdot 10^{-4}(s-4) + 8.36\cdot 10^{-6}(s-4)^2 +
1.75\cdot 10^{-7} (s-4)^3 \nonumber \\
d^2_0(s) & = &  5.09\cdot 10^{-4}(s-4) + 6.32\cdot 10^{-5}(s-4)^2 -
3.78\cdot 10^{-7} (s-4)^3 \nonumber 
\end{eqnarray}
The results are displayed in Fig. 1-3.
An inspection shows that our reconstruction of
the driving terms for these waves compares well
with the BFP fits.  

The results of the projection onto the
D- waves when evaluated in the threshold region yields the
D- wave driving term contribution to the D- wave scattering lengths.
These are presented in Table 1.

Note that BFP~\cite{BFP2} in their discussion of D- waves, redefine
the driving terms in order to ensure normal threshold behaviour.
We have not found the need to perform such a redefinition when we
work with sufficiently high precision and since we are interested
only in the scattering length and not in solving for the D- waves.
Such a redefinition yields a contribution from the resonance
of $2.7\cdot 10^{-4}$ for $a^0_2$ and $0$ for $a^2_2$ in the narrow
width approximation.  We do not use these results any further.  

We are now posed with the problem that the BFP model has not
been explicitly required to respect crossing symmetry constraints.  In order
to test the reliability of this model we now compare the results
obtained here with those from three sets of sum rules: 
Consider the Froissart-Gribov representation~\cite{MMS}:
\begin{equation} \label{FroissartGribov}
f^I_l(t)={4\over \pi (4-t)}\int_4^\infty ds' A^I_t(s',t) 
Q_l({2s'\over 4-t}-1),\, l\geq 2,
\end{equation}
where $Q_l(z)$ is the standard Neumann symbol.
The limit of this representation near threshold for $l=2$ yields 
the first set of sum rules
for the D-wave scattering lengths
and were also considered in a different context 
recently~\cite{pp}.  We find the ``Froissart-Gribov sum rules'':
\begin{eqnarray}\label{FG}
a^I_2 & = &
{1\over 15 \pi}\int_0^\infty \frac{d\nu}{(\nu+1)^3} A_t^I(\nu,4),
\end{eqnarray}
where $\nu\equiv (s'-4)/4$ is a convenient integration variable
and in the physical region denotes the square of the centre of mass
three momentum.

The second set were derived by
Wanders by writing down dispersion relations for partially symmetric
amplitudes in terms of partially symmetric homogeneous variables~\cite{wan:66}
which are reproduced below with our normalization and after
eliminating some typographical errors:
\begin{small}
\begin{eqnarray}\label{GW0}
\lefteqn{a^0_2  = 
\frac{1}{45 \pi} \int^\infty_0 d\nu \left[
            \frac{1}{(\nu+1)^3} \left[ A^0_s(\nu,0) +
	    5 A^2_s(\nu, 0)\right]
 + \frac{3(\nu^2+3\nu+1)}{\nu^2(\nu + 1)^3}A_s^1(\nu,0) \right. } \nonumber \\
\ \\
& & \left. 
+ {2\over (\nu(\nu+1))^2}\left((4\nu + 3)\frac{\partial}{\partial t}
A_s^0(\nu,0) 
 -3\nu \frac{\partial}{\partial t}
A_s^1(\nu,0) + 5\nu \frac{\partial}{\partial t}
A_s^2(\nu,0)\right)\right], \nonumber 
\end{eqnarray}
\begin{eqnarray}\label{GW2}
\lefteqn{a^0_2  =  \frac{1}{90 \pi} \int^\infty_0 d\nu \left[
            \frac{1}{(\nu+1)^3} \left[2 A^0_s(\nu,0) +
	     A^2_s(\nu, 0)\right]
 - \frac{3(\nu^2+3\nu+1)}{\nu^2(\nu + 1)^3}A_s^1(\nu,0)\right.} \nonumber \\
\ \\
      &   & \left.
+ {2\over (\nu(\nu+1))^2}\left(2\nu\frac{\partial}{\partial t}
A_s^0(\nu,0) 
 +3\nu \frac{\partial}{\partial t}
A_s^1(\nu,0) + (7\nu+6)\nu \frac{\partial}{\partial t}
A_s^2(\nu,0)\right)\right], \nonumber 
\end{eqnarray}
\end{small}
where $\frac{\partial}{\partial t} A^I_s(\nu,0)\equiv \lim_{t\to 0}
\frac{\partial}{\partial t} A^I_s(\nu,t)$.
More recently ATW~\cite{ATW} considered certain totally symmetric amplitudes
and obtained sum rules for various combinations of threshold parameters.
Combining some of these and the Wanders sum rule for 
$18 a^1_1-2 a^0_0 + 5 a^2_0$~\cite{wan:66},
we obtain for the third set (``ATW sum rules'') the following expressions:
\begin{small}
\begin{eqnarray}\label{ATW0}
\lefteqn{a^0_2  =  \frac{1}{45 \pi} \int^\infty_0 d\nu \left[
            \frac{1}{(\nu+1)^3} \left[ A^0_s(\nu,0) +
	    5 A^2_s(\nu, 0)\right]
 + \frac{3\nu^3 + 6 \nu^2 - 2 \nu -2}{(\nu(\nu + 1))^3}A_s^1(\nu,0)
\right.} \nonumber \\
\ \\
      &   & \left. +\frac{1}{(\nu(\nu + 1))^2} \left(
 \frac{18 + 20 \nu}{3}\frac{\partial}{\partial t}
A_s^0(\nu,0) 
+ 4\frac{\partial}{\partial t}
A_s^1(\nu,0) + {40\nu \over 3}\frac{\partial}{\partial t}
A_s^2(\nu,0)\right)\right], \nonumber 
\end{eqnarray}
\begin{eqnarray}\label{ATW2}
\lefteqn{a^2_2  =  \frac{1}{90 \pi} \int^\infty_0 d\nu \left[
            \frac{1}{(\nu+1)^3} \left[2 A^0_s(\nu,0) +
	     A^2_s(\nu, 0)\right]
 - \frac{3\nu^3 + 6 \nu^2 - 2 \nu -2}{(\nu(\nu + 1))^3}
A_s^1(\nu,0)
\right.} \nonumber \\
\ \\
      &   & \left. +\frac{1}{(\nu(\nu + 1))^2} \left(
\frac{16\nu}{3}\frac{\partial}{\partial t}
A_s^0(\nu,0) 
 - 4\frac{\partial}{\partial t}
A_s^1(\nu,0)
+ {36+32\nu\over 3}\frac{\partial}{\partial t}
A_s^2(\nu,0)\right)\right]. \nonumber 
\end{eqnarray}
\end{small}
The results of introducing eq.(\ref{resonance}),
(\ref{pomeron}) and (\ref{regge}) together with the crossing
relation eq.(\ref{crossingrelation}) and the numerical choices
of BFP documented in the previous section are displayed in
Tables 2, 3 and 4 for the Froissart-Gribov (eq.(\ref{FG})), Wanders
(eq.(\ref{GW0}) and (\ref{GW2})) and
ATW (eq.(\ref{ATW0}) and eq.(\ref{ATW2}))
sum rules respectively.   These may be viewed as tools that test
the extent to which the results of Table 1 are reliable since
the BFP model has not been required to satisfy crossing
constraints.  
In the event crossing constraints were to be built into the
model for medium and high energy scattering, the Roy equation
driving term contributions would have to be identical to those
obtained from any other system of sum rules.  

We see that
the entires of Table 1 are identical only to those of Table 3.
This is ostensibly due to the manner in which the Roy equations
and the Wanders' partially symmetric homogeneous variable technique
implement crossing symmetry.  

The numerical results of Table 4
are somewhat different from the above since these are based on
dispersion relations written down for totally symmetric amplitudes
in terms of totally symmetric homogeneous variables.  Nevertheless,
the results for the $\pi^0\pi^0$ combination
\begin{eqnarray*}
a_2\equiv a^0_2+2 a^2_2
\end{eqnarray*}
is identical for the entries of Table 1, 3 and 4.  
This is not unexpected
since  the Roy equations, Wanders and
ATW sum rules all involve only physical region quantities, viz.,
at the physical point $t=0$.
This is not so even for $a_2$ from
the Froissart-Gribov sum rules, which requires the evaluation
of quantities at the unphysical point $t=4$.  

Special
attention may however be paid to the Regge contribution
to $a_2$ in Tables 1-4 which are identical.  This results
from the fact that with the simplified Regge trajectory $\alpha(t)=1/2$,
the Regge contribution to the absorptive parts have no $t-$ dependence;
 a pure S- (and P-) wave contribution will automatically satisfy
crossing constraints derived from dispersion relations with two
subtractions and crossing symmetry is recovered by the
decoupling of the $I=1$ channel.

The Pomeron contributions are approximately equal
for all the 4 evaluations of $a_2$ since the absorptive parts
are dominated by S- and P- wave contributions with corrections
coming from higher waves.  The resonance contributions from
the Froissart-Gribov representation to $a_2$ differ appreciably
from the contributions computed from physical region sum rules
reflecting an unsatisfactory representation of the absorptive parts
due to the $l=2$ resonance exchange, say at the level of about
$30\%$.  Nevertheless, in toto the model yields answers for
$a^0_2$ and $a^2_2$ that always remain comparable, allowing us
to judge the BFP model as
being fair in its implementation of crossing
symmetry.

We finally remark on the numerical impact
of this work are to the results of the recent analysis of \pipi data~\cite{AB}
The resonance contributions of medium and high energy
information  which contributed $0.54\cdot 10^{-4}$
to $a^0_2$ and
$0.38\cdot 10^{-4}$ to $a^2_2$ with
the Particle Data Group
parameters for the $f_0$ were presented
there.  The numerical details with the BFP
parameters for the $f_0$ are now presented in
Tables 1-4 and also available are the Regge and Pomeron contributions
may also be included which yield $1.40\cdot 10^{-4}$ to 
$a^0_2$ and $6.82\cdot10^{-6}$ to $a^2_2$.  
In Table 5 we present our results for $a^0_2$ and $a^2_2$ for
the Roy equation fits discussed in our earlier work~\cite{AB},
explicitly accounting for the S- and P- wave contributions and
driving term contributions.
The total contribution
of the medium and high energy absorptive parts is thus seen to
be an important fraction of the central experimental value for
$a^0_2$ quoted to be $17\times 10^{-4}$ by Nagels et al.~\cite{nagels}.
When this is now completely accounted for, the results of
our recent work~\cite{AB} revise our numbers into the neighbourhood of
this number, from the neighbourhood of $15\cdot 10^{-4}$.
Our conclusions on $a^2_2$ vis a vis our earlier work are not
significantly
changed since the Regge and Pomeron contribution here is
about $25\%$ of the contribution of the resonance.

\setcounter{equation}{0}
\section{The F- wave scattering length $a^1_3$}

In the previous section we have considered the implications of
the medium and high energy \pipi scattering information to
the D- wave scattering lengths that receive contributions in
chiral perturbation theory from parameters in the Lagrangian
introduced at order $p^4$~\cite{g+l:ann}.  The F- wave scattering length
$a^1_3$ received contributions at order $p^4$ from pure
loop contributions and would be significantly modified at
the next order in chiral perturbation theory~\cite{gentwoloop,bijnens+}.  
This receives
contributions from the S- and P- wave phase shifts from dispersive
relations that might be written down for the
$I=1$, $l=3$ F- wave and also
from the medium and high energy parts.  The former were not
considered in Ref.~\cite{AB} since at that point only phenomenological
parameters at one-loop were considered.
However, we will use this opportunity to compute the medium and
high energy contributions to the F- wave scattering length, as well
as the S- and P- wave contributions.  In Table 6
we provide the S- and P- wave contributions to $a^1_3$
from the Roy equation fits for ${\rm Im}f^0_0(s')$,
${\rm Im}f^2_0(s')$ and ${\rm Im}f^1_1(s')$, employed in Ref.~\cite{AB},
upon inserting these into the $ds'$ integral in eq.(\ref{a13sp}).

Once more, the Roy equations
constributions of the medium and high energy
absorptive parts may be used by employing
the relations eq.(\ref{resonance}),
(\ref{pomeron}) and (\ref{regge})
and may be  inserted into the dispersion 
relation~(\ref{roydispersion}) and then be subsequently projected
onto the $I=1$, $l=3$ partial wave, via. eq.(\ref{projection}).  
We may once again consider the Froissart-Gribov representation
eq.(\ref{FroissartGribov}) for the F- wave and consider it in
the threshold region, which yields the sum rule:
\begin{equation}\label{a13fg}
a^1_3={1\over 70\pi} \int_0^\infty {d\nu \over (\nu+1)^4}A^1_t(\nu,4)
\end{equation}
Another sum rule for this quantity has been obtained by
Ananthanarayan, Toublan and Wanders~\cite{ATW2} from techniques of
the kind described in~\cite{wan:66,ATW} and is given below:
\begin{small}
\begin{eqnarray}\label{a13atw}
a^1_3 & = & {1\over 420 \pi}\int_0^\infty d\nu\left\{ {1\over
(\nu+1)^4}\left[2A^0(\nu,0)-5A^2(\nu,0)+{3(\nu+2)\over \nu}A^1(\nu,0)\right]
\right. \nonumber \\
& & \left.
 +{4\over \nu(\nu+1)^3}{\partial \over \partial t}(2A^0(\nu,0)-5A^2(\nu,0))
+{48\over \nu^2(\nu+1)^3}(\nu^2+4\nu+2) \right.
\\
& & \left.
{\partial \over \partial t}\left({A^1(\nu,0)\over 2t+4\nu}\right)\right\}
\nonumber .
\end{eqnarray}
\end{small}
The BFP absorptive parts for the medium and high energy parts
may be inserted into each of these sum rules and in Table 7 we
provide a compilation of the numbers of interest.
We note that the resonance contributions to the Froissart-Gribov
and ATW sum rules are identical:  this is not a numerical
coincidence; one may show that insertion of eq.(\ref{resonance})
in eq.(\ref{a13fg}) and eq.(\ref{a13atw}) yields identical
expressions.  The Pomeron yields a contribution numerically
comparable to that of the resonance only in case of the Roy equations,
reflecting a breakdown of the model.  Nevertheless, the medium and
high energy contribution is 2 orders of magnitude lower than
that of the S- and P- wave contribution.

Indeed in Ref.~\cite{ATW} it was pointed out that the sum rule
for $a^1_3$ belongs to a family of such rapidly converging ones
that it is fair to expect the contribution from the medium and
high energy tail to be small in comparison with that from the
low energy S- and P- wave contributions.  An inspection of Tables
6 and 7
indeed bears out this expectation.  It may be concluded from
Tables 6 and 7, that when Roy equation fits to \pipi scattering
data is performed with $a^0_0\in (0.19,0.21)$, the result for
$a^1_3\simeq (4.2\pm0.2)\cdot 10^{-5}$, which is compatible with the
numbers presented in Ref.~\cite{nagels}.  The one-loop chiral
prediction for this quantity is $2\cdot 10^{-5}$~\cite{g+l:ann}
and a substantial revision of this due to two-loop effects is
entirely reasonable.

\section*{Acknowledgments}  We thank the Swiss National Science
Foundation for support during the course of this work.  It is a
pleasure to thank H. Leutwyler for discussions and crucial insights.
We thank M. R. Pennington for having brought Ref.~\cite{PennProt}
to our attention.

\newpage

\noindent{\bf Figure Captions}

\bigskip

\noindent Fig. 1. Plot of $d^0_0(s,110)$ vs. s for $4\leq s\leq 60$;
solid line corresponds to our result, dashed line to the polynomial fit
BFP (I) from~\cite{BFP1} and dots to the polynomial fit 
BFP (II) from~\cite{BFP2}.  [Note
that for the $I=0$ S- wave the two polynomial fits of BFP are the same].

\bigskip

\noindent Fig. 2. As above for $d^1_1(s,110)$.

\bigskip

\noindent Fig. 3. As above for $d^2_0(s,110)$.

\newpage

\noindent{\bf Table Captions}

\bigskip

\noindent Table 1.  Contributions to $a^0_2$,
$a^2_2$ and $a_2$ from resonance, Pomeron
exchange and Regge trajectories extracted from the Roy equations.

\bigskip

\noindent Table 2.  As Table 1 but extracted from the Froissart-Gribov
representation.

\bigskip

\noindent Table 3. As Table 1 but extracted from the Wanders sum rules.

\bigskip

\noindent Table 4.  As Table 1 but extracted from the ATW sum rules.

\bigskip

\noindent Table 5 (a). Contributions to $a^0_2$ 
from the S- and P- wave Roy equation solutions of Ref.~\cite{AB},
driving term contributions and their sum, (b) As in (a) for $a^2_2$.

\noindent Table 6.  Contributions to $a^1_3$ 
from the S- and P- wave Roy equation solutions of Ref.~\cite{AB}.
[Note that the medium and high energy contributions are negligible
in comparison.]

\noindent Table 7. 
Contributions to the $I=1$, $l=3$, F- wave scattering length
$a^1_3$ from the resonance, Pomeron exchange and Regge
trajectories extracted from the Roy equations, Froissart-Gribov
sum rule and the ATW sum rule.

\newpage

$$
\begin{array}{||c|c|c|c|c||}\hline
  & {\rm Resonance}  &{\rm Pomeron} & {\rm Regge } & {\rm Total} \\ \hline
a^0_2 & 4.33\cdot 10^{-5} & 1.06 \cdot 10^{-4} & 2.84 \cdot 10^{-5} & 1.78
\cdot 10^{-4} \\
a^2_2 & 3.01\cdot 10^{-5} & 1.10\cdot 10^{-5}  & -7.58\cdot 10^{-7} & 4.03
\cdot 10^{-5} \\  
a_2 & 10.35\cdot 10^{-5} & 1.28\cdot 10^{-4} & 2.69\cdot 10^{-5} & 2.59
\cdot 10^{-4} \\ \hline
\end{array}
\medskip
$$
$$
{\rm Table\, 1}
$$

\bigskip

$$
\begin{array}{||c|c|c|c|c||}\hline
  & {\rm Resonance}  &{\rm Pomeron} & {\rm Regge } & {\rm Total} \\ \hline
a^0_2 & 3.43\cdot 10^{-5} & 1.26 \cdot 10^{-4} & 2.69 \cdot 10^{-5} & 1.88
\cdot 10^{-4} \\
a^2_2 & 3.43\cdot 10^{-5} & 0  & 0 & 3.43\cdot 10^{-5} \\ 
a_2 & 10.29\cdot 10^{-4} & 1.26\cdot 10^{-4} & 2.69\cdot 10^{-9}
& 2.57\cdot 10^{-4} \\ \hline
\end{array}
\medskip
$$
$$
{\rm Table\, 2}
$$

\bigskip

$$
\begin{array}{||c|c|c|c|c||}\hline
  & {\rm Resonance}  &{\rm Pomeron} & {\rm Regge } & {\rm Total} \\ \hline
a^0_2 & 4.33\cdot 10^{-5} & 1.06 \cdot 10^{-4} & 2.84 \cdot 10^{-5} & 1.78
\cdot 10^{-4}\\
a^2_2 & 3.01\cdot 10^{-5} & 1.10\cdot 10^{-5}  & -7.58\cdot 10^{-7} & 4.03
\cdot 10^{-5} \\ 
a_2 & 10.35\cdot 10^{-5} & 1.28\cdot 10^{-4} & 2.69\cdot 10^{-5} &
2.59 \cdot10^{-4} \\ \hline
\end{array}
\medskip
$$
$$
{\rm Table\, 3}
$$

\bigskip

$$
\begin{array}{||c|c|c|c|c||}\hline
  & {\rm Resonance}  &{\rm Pomeron} & {\rm Regge } & {\rm Total} \\ \hline
a^0_2 & 4.05\cdot 10^{-5} & 1.13 \cdot 10^{-4} & 2.78 \cdot 10^{-5} & 1.81
\cdot 10^{-4} \\
a^2_2 & 3.15\cdot 10^{-5} & 7.32\cdot 10^{-6}  & -4.96\cdot 10^{-7} & 3.83
\cdot 10^{-5} \\ 
a_2 & 10.35\cdot 10^{-5} & 1.28\cdot 10^{-4} & 2.69\cdot 10^{-5} &
2.59 \cdot10^{-4} \\ \hline
\end{array}
\medskip
$$
$$
{\rm Table\, 4}
$$

$$
\begin{array}{||c|c|c|c||}\hline
a^0_0 &\mbox{S- and P-} & \mbox{Driving term} & \mbox{Total} \\ \hline
0.19 & 14.1\cdot 10^{-4} & 1.8\cdot 10^{-4} & 15.9\cdot 10^{-4} \\
0.20 & 14.1\cdot 10^{-4} & 1.8\cdot 10^{-4} & 15.9\cdot 10^{-4} \\
0.21 & 14.2\cdot 10^{-4} & 1.8\cdot 10^{-4} & 16.0\cdot 10^{-4} \\ \hline
\end{array}
$$
$$
{\rm Table\, 5(a)}
$$

$$
\begin{array}{||c|c|c|c||}\hline
a^0_0 &\mbox{S- and P-} & \mbox{Driving term} & \mbox{Total} \\ \hline
0.19 & 0.18\cdot 10^{-4} & 0.40\cdot 10^{-4} & 0.58\cdot 10^{-4} \\
0.20 & 0.29\cdot 10^{-4} & 0.40\cdot 10^{-4} & 0.69\cdot 10^{-4} \\
0.21 & 0.41\cdot 10^{-4} & 0.40\cdot 10^{-4} & 0.81\cdot 10^{-4} \\ \hline
\end{array}
$$
$$
{\rm Table\, 5(b)}
$$

$$
\begin{array}{||c|c||}\hline
a^0_0 & a^1_3  \\ \hline
0.19 & 4.1\cdot 10^{-5} \\
0.20 & 4.2\cdot 10^{-5} \\
0.21 & 4.4\cdot 10^{-5} \\ \hline
\end{array}
$$
$$
{\rm Table\, 6}
$$

$$
\begin{array}{||c|c|c|c|c||}\hline
  & {\rm Resonance}  &{\rm Pomeron} & {\rm Regge } & {\rm Total}  \\ \hline
\mbox{Roy  equation} & 3.14\cdot 10^{-7} & 4.16\cdot 10^{-7} & 1.34\cdot 
10^{-7} & 8.64 \cdot 10^{-7} \\
{\rm Froissart-Gribov} & 3.58\cdot 10^{-7} & 0 & 1.26\cdot 10^{-7} & 4.84
\cdot 10^{-7} \\
{\rm ATW} & 3.58\cdot 10^{-7} & 1.77\cdot 10^{-9} & 1.25\cdot 10^{-7}
& 4.85 \cdot 10^{-7} \\ \hline
\end{array}
\medskip
$$
$$
{\rm Table\, 7}
$$

\newpage

\begin{figure}[h]
\epsfxsize=8cm
\centerline{\epsffile{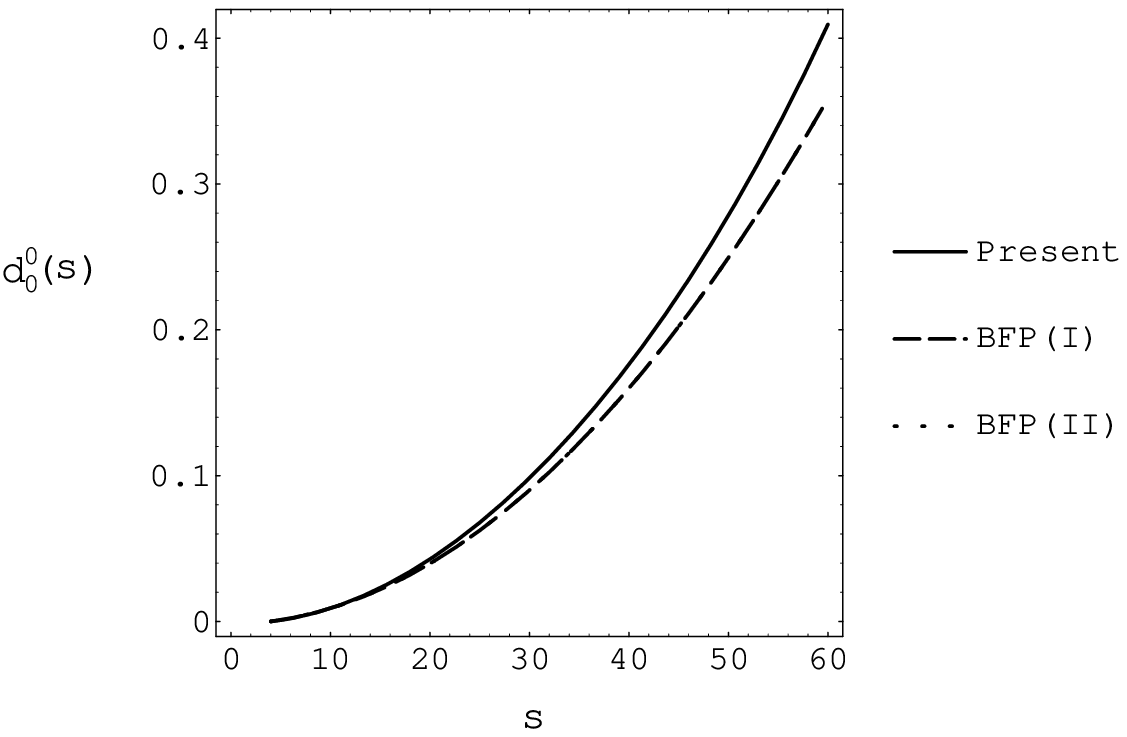}}
\centerline{Fig.~1}
\end{figure}

\begin{figure}[h]
\epsfxsize=8cm
\centerline{\epsffile{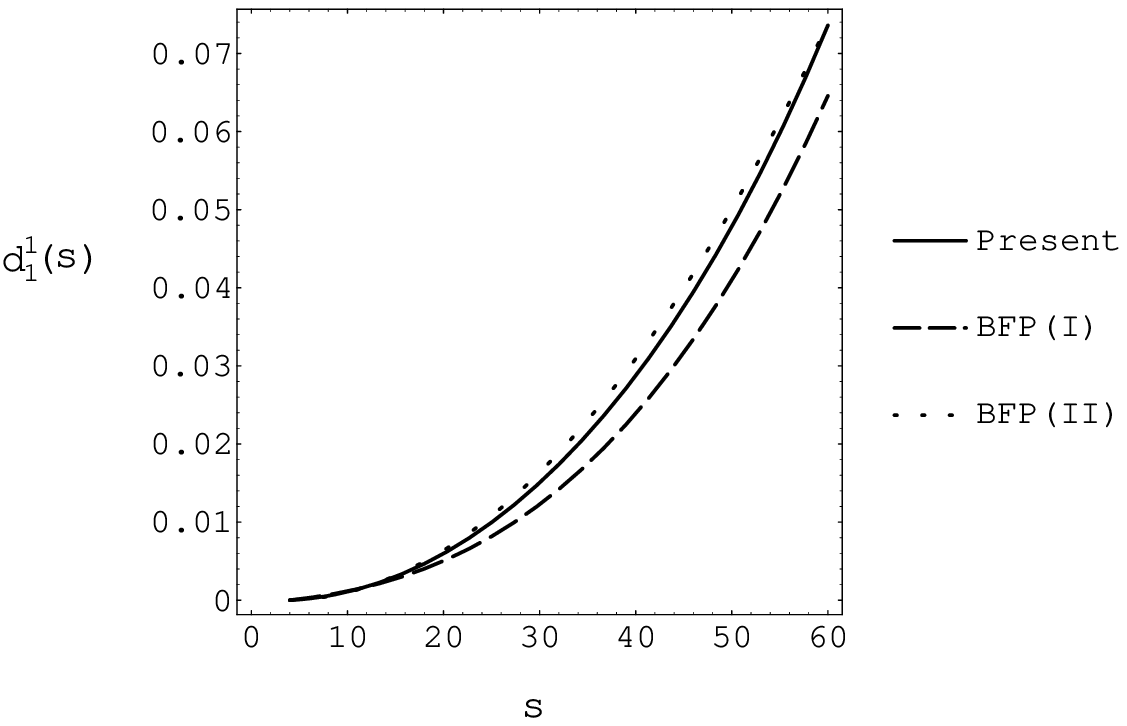}}
\centerline{Fig.~2}
\end{figure}

\begin{figure}[h]
\epsfxsize=8cm
\centerline{\epsffile{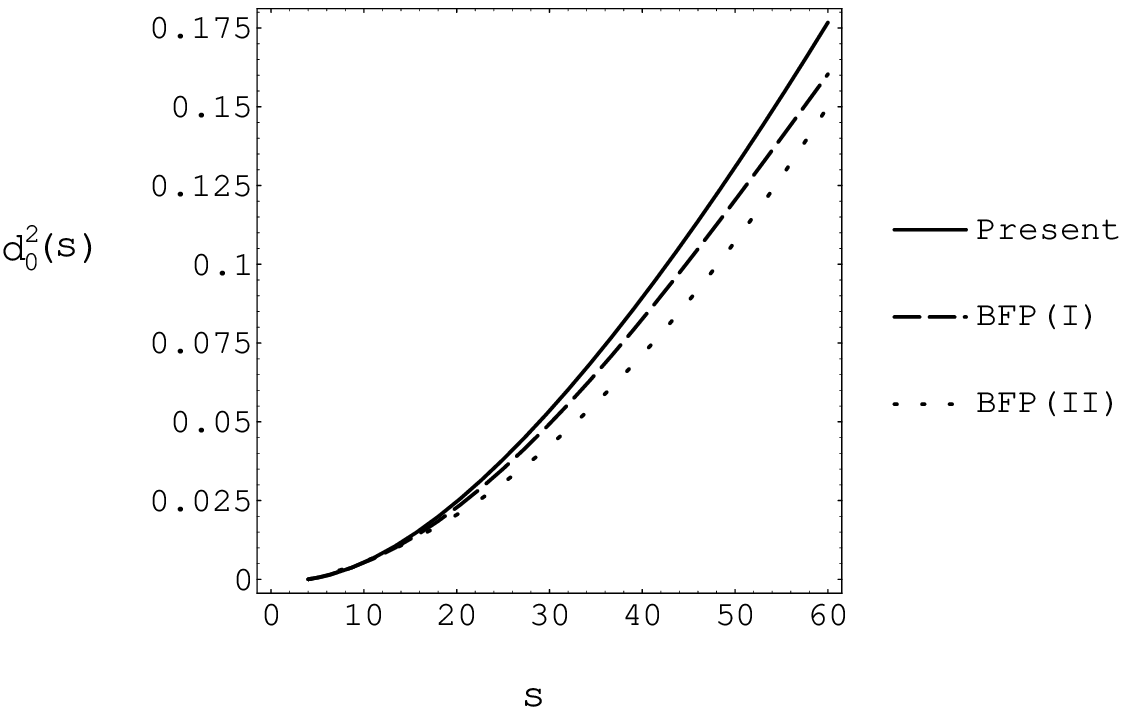}}
\centerline{Fig.~3}
\end{figure}

\end{document}